
\magnification 1200
\font\abs=cmr9
\font\ccc=cmcsc10
\def\uno{{\bf 1}}
\def\tens{\otimes}
\def\fraz#1#2{{\strut\displaystyle #1\over\displaystyle #2}}
\def\tende#1#2{\matrix{\phantom{x}\cr \longrightarrow\cr{#1\rightarrow #2
               \atop\phantom{x}}\cr}}
\def\esp#1{e^{\displaystyle#1}}
\def\ii#1{\item{$\phantom{1}[#1]~$}}
\def\jj#1{\item{$[#1]~$}}
\def\pl{\pi_{{}_L}}
\def\pr{\pi_{{}_R}}
\def\ac{a^\dagger}
\def\h{{\cal H}(1)}
\def\u{{\cal U}}
\def\uh{{\cal U}_\hbar}
\def\str{*_\hbar}
\def\fidi{\hskip5pt \vrule height4pt width4pt depth0pt}
\hsize= 15 truecm
\vsize= 22 truecm
\hoffset= 1. truecm
\voffset= 0.3 truecm

\baselineskip= 14 pt
\footline={\hss\tenrm\folio\hss} \pageno=1

\vglue 3truecm
\centerline{\bf DEFORMATION QUANTIZATION OF THE}
\centerline{\bf HEISENBERG GROUP.}
\bigskip
\bigskip
\centerline{{\it
F.Bonechi ${}^1$, R.Giachetti ${}^2$, E.Sorace ${}^1$ and M.Tarlini ${}^1$.}}
\bigskip
${}^1$Dipartimento di Fisica, Universit\`a di Firenze and INFN--Firenze,

${}^2$Dipartimento di Matematica, Universit\`a di Bologna.
\bigskip
\bigskip

{\bf Abstract.} {\abs A $*$-product compatible with the comultiplication
of the Hopf algebra of the functions on the Heisenberg group
is determined by deforming a coboundary Lie-Poisson structure defined by a
classical $r$-matrix satisfying the modified Yang-Baxter equation.
The corresponding quantum group is studied and its $R$-matrix is
explicitly calculated.}

\bigskip
\bigskip

{\bf 1. Introduction.}
The quantization of a dynamical system on a symplectic manifold was
introduced in [1,2] by deforming the pointwise multiplication of the
commutative algebra of the classical observables into a one parameter
family $\str$ of associative but not necessarily commutative
products. The parameter $\hbar$ is physically interpreted as the Planck
constant and the deformation is required to satisfy the classical limit
conditions
$$\phi\str \psi\tende \hbar 0 \phi\psi\,,~~~~~~~~~~
(\phi\str \psi - \psi\str \phi)/\hbar
\tende \hbar 0 \{\phi,\psi\}$$
for any pair of observables $\phi,\psi$.

Since its first appearance, the method has found a constantly increasing
number of applications and a special attention has been devoted to systems with
symmetry. For instance the geometric quantization, or coadjoint orbit method,
yielding the irreducible representations of nilpotent Lie groups has been
reproduced in this approach [3]; the deformation of quotient
manifolds of the Heisenberg group by appropriate lattice subgroups has been
investigated in [4] in connection with results on quantum tori; a framework
for quantizing the linear Poisson structures has been proposed in [5].
Almost all deformations are expressed in terms of formal power series in
$\hbar$ with coefficients in the algebra of the observables and
a very tiny number of $\str$--products is explicitly known, the most relevant
of which is obtained by the Weyl quantization on ${\bf R}^{2n}$. In [4,5] the
convergence of the power series is discussed and an answer
is provided in terms of Fourier transforms.

A natural question that arises in this context concerns the relationship and
the applicability of these results to quantum groups, being themselves
deformations of the algebras of the representative functions of Lie groups.
A general prescription for handling the problem and relating quantum groups to
$\str$--products is presented in [6]. Here the fundamental objects
are assumed to be a classical $r$--matrix and the corresponding coboundary
Poisson--Lie structure determined by $r$. This implies that the resulting
$\str$--product will be compatible with
the comultiplication of the Hopf algebra of the functions on the group,
differently from [4,5], where the Poisson brackets to be quantized are
required to be left--invariant for the appropriate action of the Lie group,
so that the deformed multiplication inherits the same property.

A distinction in the procedure of quantization must be made according
to whether the $r$--matrix satisfies the CYBE or the MYBE (respectively:
classical and modified Yang--Baxter equation). While a theory [7] and
some explicit results exist for
the former case [8], no deformed product, at our knowledge, has been found
in the second: applications
to quantum groups have only been made with the explicit use of a representation
and reproduce, for instance, the expression of the quantum $R$--matrix of
$SL_q(2)$ in the fundamental representation.

In this paper we shall consider the deformation quantization of the one
dimensional Heisenberg group $H(1)$. After some remarks on the MYBE for
$H(1)$ we determine the coboundary Poisson Lie structure and the brackets
to be quantized. We then look for a deformation of the pointwise multiplication
of functions on $H(1)$ and we give an explicit form for the $\str$--product
whose restriction to symplectic leaves is obviously equivalent
to Weyl quantization, but has different invariance properties.
We finally calculate the $R$--matrix for the quantized structure
thus found, obtaining the same expression as the one found
in [9,10] for the Heisenberg quantum group $H_q(1)$.
A straightforward generalization to the Heisenberg group in $n$ dimensions
is finally given and the quantum group $H_q(n)$ is determined.

\bigskip

{\bf 2. Coboundary Poisson--Lie structure.}
We denote by $a,\ac,h$ the three generators of the Heisenberg Lie algebra $\h$
that satisfy the commutation relations
$$[a,\ac]=h\,,~~~~~~~[h,\cdot\,]=0\,.\eqno(2.1)$$
Let
$$r=\lambda\,a\wedge\ac + \mu\,a\wedge h + \nu\,\ac\wedge h\ \ \in\;\wedge^2\h
  \eqno(2.2)$$
and denote by $\u=U(\h)$ the enveloping algebra of $\h$. Let us also introduce
the notation $r_{ij}$, $i=1,2,3;\,i<j$, where $r_{12}=r\tens\uno\in
\tens^3 \u$ and similarly for the other values of indices.
\medskip
{\ccc (2.3) Lemma.} {\it $r$ satisfies the MYBE.
If $\lambda=0$, $r$ satisfies the CYBE.}

{\sl Proof.} Defining
$$B=[r_{12},r_{13}]+[r_{12},r_{23}]+[r_{13},r_{23}]$$
a direct calculation shows that $Ad_\kappa B=0$ for any $\kappa\in\h$, where
$Ad$ denotes the diagonal adjoint action of $\h$ on $\tens^3 \u$. Also
$B=0$ for $\lambda=0$.\fidi
\medskip

Therefore, according to the usual definition, any element of $\wedge^2\h$
is a classical $r$--matrix.

Let us define on $H(1)$ the coordinates $(\beta,\delta,\alpha)$ such that
for $g\in H(1)$ and for $(z,y,x)=(\beta(g),\delta(g),\alpha(g))$ we have
$$g=\esp{zh}\,\esp{y\ac}\,\esp{xa}\,\equiv\,
  (z,y,x)\,.$$
The composition law
$$ (z',y',x')\cdot(z,y,x)=(z'+z+x'y,\,y'+y,\,x'+x)$$
induces then the comultiplication
$$\eqalign{
  \Delta\delta=1\tens\delta & +\delta\tens 1\,,~~~~~~~~~~
  \Delta\alpha=1 \tens\alpha+ \alpha\tens 1\,,\cr
{} & \Delta\beta=1 \tens\beta+\beta\tens 1+\alpha\tens\delta\,.\cr}$$

\medskip
{\ccc (2.4) Proposition.} {\it The coboundary Lie--Poisson structure
associated to an element $r$ as in $(2.2)$ has the following brackets:}
$$\{\delta,\beta\}=\lambda\delta\,,~~~~~~~
  \{\delta,\alpha\}=0\,,~~~~~~~\{\alpha,\beta\}=\lambda\alpha\,.$$

{\sl Proof}. If $ad$ denotes the adjoint representation of $H(1)$ on $\h$,
for $g=(z,y,x)$ we have
$$ad_{g}\,a=a-y h\,,~~~~~~~ad_{g}\,\ac=\ac+x h\,,~~~~~~~
  ad_{g}\,h=h\,.$$
Since the right invariant fields on $H(1)$ are
$$X_{h}=\partial_{\beta}\,,~~~~~~~
  X_{\ac}=\partial_{\delta}\,,~~~~~~~
  X_{a}=\delta\partial_{\beta}+\partial_{\alpha}\,,\eqno(2.5)$$
the brackets $\{\,,\,\}=\eta_{uv}X_uX_v$, ($u,v=a,\ac,h$), determined by the
coboundary
$$\eta=ad\;r-r=\lambda(a\wedge \alpha\,h +\ac\wedge \delta\,h)\,
:\,H(1)\longrightarrow \wedge^2\h\,,\eqno(2.6)$$
are of the stated form.\fidi
\medskip

{\ccc (2.7) Remark}.
It can be observed that the Poisson brackets are not vanishing if and only if
$\lambda$ is not vanishing and in this case they differ by a multiplicative
factor, so that the coboundary Lie--Poisson structure is essentially
unique. In what follows, without loss of generality, we shall thus fix
$\lambda=1/2$, $\mu=\nu=0$, so that $r={1\over 2} a\wedge\ac$.

\bigskip

{\bf 3. The $\str$--product.}
Let $\uh$ be the algebra obtained from $\u$ by extending the field of
coefficients to the ring of formal power series in $\hbar$. Let then
$\pl$ and $\pr$ respectively be the representation and the
anti-representation of the Lie algebra $\h$ by left and
right invariant vector fields on $H(1)$ and use the same notation for
the extension of the representation that maps $\uh$ into the left and right
invariant differential operators on $H(1)$. Consider an element
$F \in \tens^2\,\uh$ of the form
$$F=\uno\tens\uno+{\hbar\over2}\; r + \cdots \ ,\eqno(3.1)$$
for which
$$(\uno\tens\epsilon) F=(\epsilon\tens\uno) F=\uno\tens\uno\ .$$
Where $\uno$ is the unity, $\epsilon$ the counit
of $\uh$ and $r$ is the element specified in (2.7). Moreover $F$
is invertible in $\tens^2\,\uh$: letting
$\pl^{\tens}=\pl\tens\pl$, $\pr^{\tens}=\pr\tens\pr$, we can define
$$\widetilde{F}=\pl^{\tens}F\,,~~~~~~~~~~
{\widetilde{F}}'=\pr^{\tens}(F^{-1})$$
and a composition law [6]
$$\str=m\circ \widetilde{F}\circ {\widetilde{F}}'\eqno(3.2)$$
on the differentiable functions on $H(1)$, $m$ denoting the
multiplication of functions. According to the standard
use, the law (3.2) will be called $\str$--product.
If $\phi,\psi\in C^{\infty}(H(1))$, then the properties
$$\eqalign{i)&\ \ \ \phi\str 1 = 1\str\phi = \phi\ ,\cr
ii)&\ \ \ \{\phi,\psi\} = \lim_{\hbar\rightarrow 0}\fraz{1}{\hbar}(\phi\str\psi
- \psi\str\phi)\ ,\cr
iii)&\ \ \ \Delta(\phi\str\psi) = \Delta(\phi)\str\Delta(\psi)\ ,\cr}$$
are easily verified. To deal with the associativity of the $\str$--product
we recall the following result:

\medskip
{\ccc (3.3) Proposition} [6].
{\it Assume that an element $F$ as given in $(3.1)$
satisfies the equation
$$(\Delta\tens \uno)F\;(F\tens \uno)\ =\ \chi\ (\uno\tens\Delta)F\;
(\uno\tens F)\ ,
\eqno(3.4)$$
where $\chi\in\tens^3\,\u$ is invariant, i.e. $Ad_\kappa\chi= 0$ for
any $\kappa\in\h$. Then the product $\str$ as in $(3.2)$ is associative.}\fidi
\medskip

Therefore we shall prove the associativity by solving equation (3.4).
Let ${\widehat \uh}=\uh\tens_Z Q(Z)$ be the localization of $\uh$
with respect to its center $Z$, $Q(Z)$ being the field
of fractions of $Z$. Let then ${\bf x}=h\tens \uno\ , \ \
{\bf y}=\uno\tens h\ ,
\ \ \uno^\tens =\uno\tens\uno$ and define the elements
$\theta_1,\theta_2\in {\tens^2\,\widehat\uh}$ by the relations
$$\tan{\theta_1}=\sqrt{\fraz{{\bf y}}{{\bf x}}}\ ,\ \ \ ~~~~~
\tan{\theta_2}=\sqrt{\fraz{\uno^\tens-e^{-\hbar {\bf y}}}
{e^{\hbar {\bf x}}-\uno^\tens}}\,.$$

\medskip

{\ccc (3.5) Proposition.} {\it If $\theta=\theta_1-\theta_2$ and
$\rho=2\theta/\sqrt{{\bf x}{\bf y}}$, then $\rho\in\tens^2\,\uh$
and the element $$F\;=\;e^{\displaystyle \;\rho\,r}$$
solves $(3.4)$.}

{\sl Proof.} For the first statement it is sufficient to
observe that $\ \tan{\theta_2} = \sqrt{{\bf y}/{\bf x}}\,(1-\hbar({\bf x}+
{\bf y})\,\Theta(\hbar,{\bf x},{\bf y}))\ $,
where $\Theta$ is a power series in $\hbar$, ${\bf x}$, ${\bf y}$.
We then obtain $\theta = \theta_1-\theta_2 = \arctan \{\,\sqrt{{\bf x}{\bf
y}}\,\hbar\Theta/(1-{\bf y}\,\hbar \Theta)\,\}$: thus
$\rho=2\theta/\sqrt{{\bf x}{\bf y}}\,\in\,\tens^2\,\uh$. The first terms of
the expansion of $\rho$ in powers of $\hbar$ are
$$\rho = \fraz{\hbar}2 - \fraz{\hbar^2({\bf x}-{\bf y})}{48} +
O(\hbar^3)\,.$$
To prove the second statement, introduce the notation
$F_{12,3}=(\Delta\tens\uno)F$, $~F_{12}=F\tens\uno$, $~F_{23}=\uno\tens F$,
$~F_{1,23}=(\uno\tens\Delta)F$. Equation (3.4) becomes
$$\chi\ =\ F_{12,3}\,F_{12}\,F_{23}^{-1}\,F_{1,23}^{-1}\ .$$
and the invariance of $\chi$ reads
$$F_{23}^{-1}\,F_{1,23}^{-1}\,(\uno\tens\Delta)\Delta \kappa\,
F_{1,23}\,F_{23} =
F_{12}^{-1}\,F_{12,3}^{-1}\,(\uno\tens\Delta)\Delta \kappa\,
F_{12,3}\,F_{12}\ ,\eqno{(3.6)}$$
with $\kappa=a,\ac$.
Defining  $f,\; g$ by
$$f=\cos\theta+\sqrt{\fraz{{\bf y}}{{\bf x}}} \sin\theta\,\,,~~~~~~~
g=\cos\theta-\sqrt{\fraz{{\bf x}}{{\bf y}}} \sin\theta$$
and using for $f$ and $g$ the same notation as for $F$, equation (3.6)
is equivalent to
$$\eqalign{
(f_{1,23}-f_{12}\,f_{12,3})\;&\kappa\tens\uno\tens\uno\; +
(g_{12}\,f_{12,3}-f_{23}\,g_{1,23})\;\uno\tens\kappa\tens\uno\; +\cr
{} &(g_{23}\,g_{1,23}-g_{12,3})\;\uno\tens\uno\tens\kappa =0\,.\cr}
\eqno(3.7)$$
With our definition of $\theta$ we can write
$$\eqalign{
{}&f=\esp{\hbar{\bf y}/4}~
\sqrt{\fraz{{\bf x}+{\bf y}}{\sinh(\hbar({\bf x}+{\bf y})/2)}}~
\sqrt{\fraz{\sinh(\hbar{\bf x}/2)}{{\bf x}}}\;,\cr
{}&g=\esp{-\hbar{\bf x}/4}~
\sqrt{\fraz{{\bf x}+{\bf y}}{\sinh(\hbar({\bf x}+{\bf y})/2)}}~
\sqrt{\fraz{\sinh(\hbar{\bf y}/2)}{{\bf y}}}\;,\cr}$$
from which equation (3.7) is verified.\fidi
\medskip
{\ccc (3.8) Corollary.} {\it
The explicit expression for the $\str$--product is as follows:}
$$\phi\str\psi\, =\, m\circ\exp\{\pr^{\tens}(\rho\eta)\}\,(\phi\otimes\psi)\,$$
{\it where $\rho$ is given in $(3.5)$ and $\eta$ in $(2.6)$.}

{\sl Proof.} Indeed
$$\eqalign{\widetilde{F}\circ {\widetilde{F}}'&=
\exp\{(\pl^{\tens}(\rho r)\}\ \exp\{-
\pr^{\tens}(\rho r)\}\cr
{}&=\exp\{\pr^{\tens}
(\rho\;(ad\; r - r))\}\;=\;
\exp\{\pr^{\tens}(\rho\eta)\}\ .~~\fidi\cr}$$
\medskip

{\ccc (3.9) Remark.}
According to (2.4), the relation $\alpha/\delta=cost.$ defines the
symplectic leaves of $H(1)$, on which the local Darboux coordinates are
$p=\log \alpha$ and $q=2 \beta$. The expression given in (3.8) can be
compared with the $*_{{}_W}$--product obtained by the Weyl quantization that
can be defined on any symplectic leaf: as expected, the latter shows to be
not compatible with comultiplication ({\it e.g.} compare
$\beta\,*_{{}_W}\,\alpha\beta^2$). This fact was already noticed in [6]
in connection with the quantization of $SL(2)$.

\bigskip

{\bf 4. $R$--matrix and the quantum group $H_q(1)$.} In this last section
we shall determine the $R$-matrix from the given quantization and we
shall compare the result with what was found in [9,10].
By a direct calculation the following lemma is easily proved.
\medskip
{\ccc (4.1) Lemma.} {\it Let
$$n=\fraz{\ac a}{h}\,,~~~~~~~~~~\xi=h\tens n + n\tens h\,.$$
The element $t\in\tens^2\,\widehat\uh$ given by
$$t=\fraz12\,(a\tens\ac+\ac\tens a -\xi\,)$$
is invariant and satisfies the equation
$$B=[t_{23},t_{13}]\,,$$
where $B$ is given in $(2.3)$. \fidi}
\medskip
Denote by $\sigma$ the flip isomorphism of $\tens^2\,\widehat\uh$,
$\sigma(a\tens b)=(b\tens a)$. The main result of this section is formulated
as follows.
\medskip
{\ccc (4.2) Proposition.} {\it The element of
$\tens^2\,\widehat\uh$
$$R=(\sigma\circ F)^{-1}\,\esp{\hbar t}\,F\ ,\eqno(4.3)$$
with $F$ as in $(3.5)$ satisfies the QYBE $($quantum
Yang--Baxter equation$\,)$ and coincides with the quantum R--matrix of
$U_q(\h)$ $[9]$.}
\medskip
The proof of this proposition is a consequence of the following two lemmas
that stem from the observation that the fundamental objects in terms of
which $F$ and $t$ are built are $a\tens\ac$, $\ac\tens a$ and $\xi$.
\medskip
{\ccc (4.4) Lemma.} {\it The elements of $\tens^2\,\widehat\uh$
given by
$$\eqalign{
   j_+=\fraz{a\tens\ac}{\sqrt{{\bf xy}}}\,,~~~~~
  & j_-=\fraz{\ac\tens a}{\sqrt{{\bf xy}}}\,,~~~~~
   j_3=\fraz{h\wedge\ac a}{2{\bf xy}}\,,\cr
   {}&k= \fraz{\xi}{\gamma} -j_3\,,\cr}$$
with $\gamma={\bf x}-{\bf y}$, generate an $u(2)$ Lie algebra.}

{\sl Proof.} Indeed it is immediate to verify that $j_{\pm},j_3$ satisfy
the $su(2)$ commutation relations, while $k$ is the central generator of
$u(1)$.\fidi
\medskip
{\ccc (4.5) Lemma} {\it The elements
$$ A= \sqrt{\fraz{2\sinh(wh/2)}{wh}}\,a\,,~~~~~
   A^\dagger=\sqrt{\fraz{2\sinh(wh/2)}{wh}}\,\ac\,,~~~~~H=h\,,\eqno(4.6)$$
satisfy the commutation relations of the generators of $U_q(\h)$ $[9]$,
namely}
$$[A,A^\dagger]=\fraz 2w\,\sinh(wH/2)\,,~~~~~~~[H,\,\cdot\,]=0\,.~~\fidi$$
\medskip
{\sl Proof of $(4.2)$.} In terms of the $u(2)$ generators, (4.3) reads
$$R=\esp{(\sigma\circ\theta)\;(j_+-j_-)}\,
\esp{\hbar/2(\sqrt{{\bf xy}}(j_{+}+j_{-})-\gamma k-\gamma j_3)}\,
\esp{\theta\;(j_+-j_-)}\,.$$
Using the composition law of the group $U(2)$, the last expression
is transformed into
$$R=\esp{-\hbar(\gamma k +\gamma j_3)/2}\,\exp\{2\esp{\hbar\gamma/4}\,
\sqrt{\sinh(\hbar{\bf x}/2)\sinh(\hbar{\bf y}/2)}\,j_+\}\,.\eqno(4.7)$$
{}From (4.4) and (4.6), letting
$$N=\fraz{wA^\dagger A}{2\sinh(wH/2)}\,,~~~~~~~\Omega=H\tens N+N\tens H\,,
\eqno(4.8)$$
equation (4.7) with $\hbar=w$ becomes
$$ R=\esp{-w\Omega/2} \exp\biggl\{w\esp{wH/4}A\tens \esp{-wH/4}A^{\dagger}
\biggr\}\,.\eqno(4.9)$$
This expression coincides with the $R$ matrix of $U_q(\h)$ found in [9],
where it is explicitly shown that $R$ satisfies the QYBE.\fidi
\medskip
{\ccc (4.10) Remarks.} $(i)$ We observe that $F$ as given in (3.5)
deforms the comultiplication $\Delta$ of $\u$ to the comultiplication
$\Delta_q$ of $U_q(\h)$, namely
$$\Delta_q(\kappa) = F^{-1}\Delta(\kappa)F\,,~~~~~~~\kappa\in\h\,,$$
as expected.
\smallskip
$(ii)$ A final remark concerns a straightforward generalization to the
deformation of the Heisenberg group $H(n)$ in $n$ degrees of freedom,
with algebra ${\cal H}(n)$
generated by $h,\ac_i,a_i$ $(i=1,...,n)$, relations
$$[a_i,\ac_j]=\delta_{ij}\,h\,,~~~~~~~[h,\cdot\,]=0$$
and corresponding group coordinates $\beta,\delta_i,\alpha_i$.
The classical $r$ matrix
$${}\phantom{\lambda_i\in{\bf R}}~~~~~
r\,=\,\sum\limits_{i=1}^n\,\lambda_i r_i\,,~~~~~~~~~~~r_i=\fraz 12 a_i\wedge
\ac_i\,\in\wedge^2{\cal H}(n)\,;~~~\lambda_i\in{\bf R}\,,$$
gives $H(n)$ a Lie-Poisson structure with coboundary
$$\eta=\sum\limits_{i=1}^n\,\lambda_i\,(a_i\wedge\alpha_i\, h +\ac_i
\wedge \delta_i\, h)\,.$$
Let $\rho_i$ be defined as in (3.5) with the substitution of $\hbar$ by
$\lambda_i\hbar$. Then
$$F=\prod\limits_{i=1}^n\,\esp{\,\rho_ir_i}$$
defines a $\str$--product on $H(n)$. The corresponding quantum group
$H_q(n)=U_q({\cal H}(n))$ is generated by $H,A_i,A^\dagger_i$
defined for each $i=1,...,n$ as in (4.6) with $w_i=\lambda_i\hbar$ replacing
$w$. The $R$-matrix of $H_q(n)$ is $\,R=\prod_i\,R_i\,$ where the $R_i$
are given by (4.8) and (4.9)
with the above substitutions.

\bigskip
\bigskip

\centerline{{\bf References.}}

\bigskip

\ii 1 J. Vey, Comment. Math. Helv., {\bf 50}, 421 (1975).
\smallskip
\ii 2 F. Bayen, M. Flato, C. Fronsdal, A. Lichnerowicz and D. Sternheimer,
      Ann. Phys., {\bf 110}, 61 and 111 (1978).
\ii 3 D. Arnal, Pacific J. Math., {\bf 114}, 285 (1984).
\smallskip
\ii 4 M.A. Rieffel, Comm. Math. Phys., {\bf 122}, 531 (1989).
\smallskip
\ii 5 M.A. Rieffel, American J. Math., {\bf 112}, 657 (1990).
\smallskip
\ii 6 L.A. Takhtajan, {\it Lectures on Quantum Groups}, in Introduction to
     Quantum Group and integrable Massive Models of Quantum Field Theory,
      Nankai Lectures in Mathematical Physics, 69, (World Scientific, 1990).
\smallskip
\ii 7 V.G. Drinfeld, Sov. Math. Dokl., {\bf 28}, 667 (1983).
\smallskip
\ii 8 Ch. Ohn, Lett. Math. Phys., {\bf 25}, 85 (1992).
\smallskip
\ii 9 E. Celeghini, R. Giachetti, E. Sorace and M. Tarlini, J. Math. Phys.,
      {\bf 32}, 1155 (1991).
\smallskip
\jj {10} E. Celeghini, R. Giachetti, E. Sorace and M. Tarlini,
      ``{\it Contractions of quantum groups}'', in Lecture Notes in Mathematics
      n. 1510, 221, (Springer--Verlag, 1992).

\bye